\documentclass[final,5p,times,twocolumn]{elsarticle}

\usepackage{graphics,graphicx}
\usepackage{dcolumn}
\usepackage{bm}
\usepackage{epsfig}
\usepackage[usenames]{color}
\usepackage{hyperref} 
\usepackage{mathbbol}
\usepackage{epstopdf}
\usepackage{simplewick}
\usepackage[utf8]{inputenc} 
\usepackage{slashed}
\usepackage{stackrel}
\usepackage{mathrsfs}
\usepackage{xcolor}
\usepackage{cancel}
\usepackage{mathtools}
\usepackage{amsmath,amssymb,amsfonts,mathrsfs}
\usepackage[most]{tcolorbox}

\usepackage[utf8]{inputenc}
\usepackage{tikz, pgfplots}
\usepackage[compat=1.1.0]{tikz-feynman}
\usepackage{subcaption}

\usepackage{physics}
\usepackage{subcaption}

\usepackage{ragged2e}
\DeclareCaptionJustification{justified}{\justifying}

\usepackage[normalem]{ulem}

\def\eq#1{{Eq.~(\ref{#1})}}
\def\fig#1{{Fig.~\ref{#1}}}
\def\sec#1{{Sec.~\ref{#1}}}
\newcommand{\ben}{\begin{eqnarray*}}
\newcommand{\een}{\end{eqnarray*}}
\newcommand{\un}[1]{\underline{#1}}

\newcommand{\pd}{\partial}

\newcommand{\thalf}{\tfrac{1}{2}}

\newcommand{\as}{\alpha_s}

\makeatletter
\DeclareRobustCommand{\cev}[1]{%
  {\mathpalette\do@cev{#1}}%
}
\newcommand{\do@cev}[2]{%
  \vbox{\offinterlineskip
    \sbox\z@{$\m@th#1 x$}%
    \ialign{##\cr
      \hidewidth\reflectbox{$\m@th#1\vec{}\mkern4mu$}\hidewidth\cr
      \noalign{\kern-\ht\z@}
      $\m@th#1#2$\cr
    }%
  }%
}
\makeatother

\begin{document}

\title{Unpolarized GPDs at small $x$ and non-zero skewness}

\author[OSU]{Yuri V. Kovchegov} 
         \ead{kovchegov.1@osu.edu}
         \address[OSU]{Department of Physics, The Ohio State
           University, Columbus, OH 43210, USA}

\author[CNF,ODU,Jlab,Temple]{M. Gabriel Santiago}
        \ead{melvin.santiago@temple.edu}
        \address[CNF]{Center for Nuclear Femtography, SURA, 12000 Jefferson Avenue Newport News, VA 23606}
        \address[ODU]{Department of Physics, Old Dominion University, Norfolk, VA 23529}
        \address[Jlab]{Jefferson Lab, Newport News, VA 23606}
        \address[Temple]{Department of Physics, Temple University, Philadelphia, PA 19122}

\author[OSU]{Huachen Sun} 
         \ead{sun.2885@osu.edu}

\begin{abstract}
We study the small-$x$ asymptotics of unpolarized generalized parton distributions (GPDs) and generalized transverse momentum distributions (GTMDs). Unlike the previous works in the literature, we consider the case of non-zero (but small) skewness while allowing for non-linear contributions to the evolution equations. We first show that unpolarized GPDs and GTMDs at small $x$ are related to the eikonal dipole amplitude $N$, whose small-$x$ evolution is given by the BK/JIMWLK evolution equations, and to the odderon amplitude $\cal O$, whose evolution is also known in the literature. We then show that the effect of non-zero skewness $\xi \neq 0$ is to modify the value of the evolution parameter (rapidity) in the arguments for the dipole amplitudes $N$ and $\cal O$ from $Y = \ln (1/x)$ to $Y = \ln \min \left\{ 1/|x| , 1/|\xi| \right\}$.
\end{abstract}

\maketitle

\tableofcontents

\section{Introduction}

Understanding the non-perturbative structure of hadrons as encoded in their quark and gluon degrees of freedom is a major unsolved problem in quantum chromodynamics (QCD). One of the main tools we have for studying the three-dimensional structure is generalized parton distribution functions (GPDs) \cite{Muller:1994ses,Ji:1996ek,Radyushkin:1996nd}. They generalize the one-dimensional parton distribution functions (PDFs), leveraging momentum transfer to add two more dimensions of information to the one-dimensional momentum information encoded in PDFs. GPDs are a rich source of information on the hadronic structure, encoding various properties of the bulk hadron state, including the distributions of mass and angular momentum \cite{Ji:1996ek,Ji:1994av}, the mechanical properties of the hadron \cite{Polyakov:2002wz,Polyakov:2002yz,Burkert:2018bqq,Kumericki:2019ddg,Ji:2025qax}, and the spatial distributions of partons within hadrons \cite{Burkardt:2000za,Burkardt:2002hr,Ji:2003ak,Belitsky:2003nz}. In order to study GPDs one must measure exclusive scattering processes like Deeply Virtual Compton Scattering (DVCS) and Deeply Virtual Meson Production (DVMP), where one can apply collinear factorization to the amplitude in terms of a hard scattering subprocess and some non-perturbative hadronic structures including GPDs \cite{Ji:1996nm,Radyushkin:1996ru,Collins:1996fb,dHose:2016mda,Kumericki:2016ehc,Favart:2015umi}.  The high-energy limit of these processes, as was probed in experiments at the Hadron-Electron Ring Accelerator (HERA)~\cite{H1:2005dtp,ZEUS:2007iet, H1:2009cml, H1:2009wnw} and will be further studied at the upcoming Electron Ion Collider \cite{Accardi:2012qut,Boer:2011fh,Proceedings:2020eah,AbdulKhalek:2021gbh}, is sensitive to the GPDs in the regime where the hadron is dominated by the gluons with small (average) longitudinal momentum fraction~$x$. This same regime can be studied within the small-$x$/saturation formalism \cite{Gribov:1984tu,Iancu:2003xm,Weigert:2005us,JalilianMarian:2005jf,Gelis:2010nm,Albacete:2014fwa,Kovchegov:2012mbw,Morreale:2021pnn,Wallon:2023asa}, where instead of GPDs the non-perturbative content of the scattering process is encoded in the color dipole scattering amplitude. The conventional approach for calculating in the small-$x$ formalism takes the longitudinal momentum transfer (called skewness in the collinear GPD framework) in a process to be effectively zero \cite{Kovchegov:1999ji,Kovner:2001vi,Hentschinski:2005er,Kovner:2006ge,Hatta:2006hs,Kowalski:2006hc,Rezaeian:2012ji,Hatta:2017cte,Mantysaari:2021ryb,Mantysaari:2022kdm,Hatta:2022bxn}, while transverse momentum transfer can be encoded in the dependence of the dipole scattering amplitude on the impact parameter of the color dipole with respect to the hadron target. This neglects some of the information encoded in collinear GPDs and necessitates the use of significant phenomenological corrections to account for the effects of non-zero skewness in the data analyses \cite{Kowalski:2006hc,Toll:2012mb,Mantysaari:2016jaz}. We note here the recent work \cite{Boussarie:2023xun} aimed at constructing a generalized formalism for exclusive processes which coincides with the collinear and small-$x$ frameworks in the corresponding limits, as well as the recent work \cite{Bhattacharya:2025fnz} on calculating generalized transverse momentum distributions (GTMDs) and their various projections (GPDs, PDFs, etc.) in the shock wave formalism.

In this work, we demonstrate how non-zero skewness enters the small-$x$ formalism, while working in the leading logarithmic approximation (LLA) of the evolution equations which govern the energy dependence of the dipole scattering amplitude. The LLA is defined as resumming powers of $\as \, \ln (1/x)$, with $\as = g_s^2/(4 \pi)$ the QCD coupling. A similar conclusion for the non-forward small-$x$ evolution was derived in \cite{Bartels:1981jh} for the linear Balitsky-Fadin-Kuraev-Lipatov (BFKL) \cite{Kuraev:1977fs,Balitsky:1978ic} evolution equation. Here we account for non-zero skewness in the context of the saturation framework, obtaining a prescription for its inclusion into the dipole scattering amplitude(s) obtained by solving the full non-linear evolution equations. The structure of the paper is as follows: in \sec{sec:gpds} we show how to simplify the operator definitions of GPDs for quarks and gluons at small $x$ and small skewness and relate them to the dipole scattering amplitude. Our derivation also yields relations between the dipole scattering amplitude and the quark and gluon GTMDs \cite{Vanderhaeghen:1999xj,Martin:1999wb,Khoze:2000cy,Martin:2001ms,Ji:2003ak,Belitsky:2003nz,Diehl:2007hd,Goloskokov:2007nt,Meissner:2009ww}, which encode both the impact parameter and transverse momentum information for the distributions of quarks and gluons in hadrons. We restrict ourselves to the case of unpolarized quarks and gluons in an unpolarized hadron. In \sec{sec:skew} we show how to account for skewness in the evolution diagrams for the dipole scattering amplitude, obtaining a prescription for evolving the dipole amplitude to a rapidity given by the maximum of the (absolute values of the) longitudinal momentum fraction variable $x$ and the skewness $\xi$. We show that this prescription holds even for the diagrams contributing to the non-linear evolution. Next, in \sec{sec:obs} we study the effects of non-zero skewness, included via our prescription, on GPD-sensitive high-energy processes and on the ratio of the GPD to the PDF at the same longitudinal momentum fraction. Finally, in \sec{sec:conc} we summarize our findings and discuss further lines of study on the effects of skewness in the small-$x$/saturation framework. 

Throughout this work we make use of light cone coordinates defined as $u^{\mu} = (u^+, u^-, \un{u})$ with $u^{\pm} = (u^0 \pm u^3) / \sqrt{2}$ and the inner product $u \vdot v = u^+ v^- + u^- v^+ - \un{u} \vdot \un{v}$. The transverse vectors are written as $\un{u} = (u^1, u^2)$, while the transverse integration measures are given by $\dd[2]{u}_{\perp}$. The magnitude of a transverse vector is denoted by $u_T = |\un u|$. Throughout this work, we assume that the hadron is moving mainly in the $x^+$ direction.

%%%%%%%%%%%%%%%%%%%%%%%%%%%%%%%%%%%%%%

\section{Unpolarized GPDs and GTMDs at small $x$}\label{sec:gpds}

The leading-twist GPDs are defined as the matrix elements of a non-local product of two gluon field strength operators $F_a^{\mu \nu} = \partial^{\mu} A^{a \, \nu} - \partial^{\nu} A^{a \, \mu} + g_s \, f^{abc} \, A^{b \, \mu} \, A^{c \, \nu}$,
\begin{align}
    \int\limits_{-\infty}^\infty \frac{\dd{z}^-}{2\pi} e^{ix \bar{P}^+ z^-} \!\! \bra{P', S'} F_b^{+i} \left(-\tfrac{z}{2} \right) U^{ba}_{\un 0} \left[-\tfrac{z^-}{2}, \tfrac{z^-}{2} \right] F_a^{+j} \left(\tfrac{z}{2} \right) \ket{P, S} ,
\end{align}
in the gluon case and as an analogous product of field operators
\begin{align}
    \int\limits_{-\infty}^\infty \frac{\dd{z}^-}{2\pi} e^{ix \bar{P}^+ z^-} \bra{P', S'} \, \bar{\psi} \left(-\tfrac{z}{2} \right) V_{\un 0} \left[-\tfrac{z^-}{2}, \tfrac{z^-}{2} \right] \, \Gamma \, \psi \left(\tfrac{z}{2} \right) \ket{P, S} ,
\end{align}
in the quark case. Here, $z^\mu = (0^+, z^-, {\un 0})$, $i,j = 1,2$, $\bar{P}^+ \equiv (P'^+ + P^+)/2$, $\Gamma$ is a matrix in spinor space, and $U_{\un 0}^{ba} [-z^-/2, z^-/2]$ is a Wilson line in the adjoint representation, defined as
\begin{align}
    U_{\un x} [x_f^- , x_i^-] \equiv \mathcal{P} e^{i g_s \int\limits_{x_i^-}^{x_f^-} \dd{x}^{-} A^{a \, +} (0^+, x^-, {\un x} ) \, T^a } ,
\end{align}
with the SU($N_c$) generator matrices $(T^a)_{bc} = -i f_{abc}$ in the adjoint representation and $\mathcal{P}$ the path-ordering operator. ($N_c$ is the number of quark colors.) We also have $V_{\un 0} [-z^-/2,z^-/2]$, a Wilson line in the fundamental representation,
\begin{align}
    V_{\un x} [x_f^- , x_i^-] \equiv \mathcal{P} e^{i g_s \int\limits_{x_i^-}^{x_f^-} \dd{x}^{-} A^{a \, +} (0^+, x^-, {\un x} ) \,  t^a} ,
\end{align}
with generator matrices $t^a$ in the fundamental representation. We denote the average fraction of the longitudinal momentum of the hadron which is carried by the quark or gluon as $x$. 

As established in the pioneering work \cite{Hatta:2022bxn} (see also the more recent \cite{Bhattacharya:2025fnz}), and by analogy to the case of PDFs \cite{Mueller:1999wm, Kovchegov:2015zha, Kotko:2015ura, Marquet:2016cgx, Hatta:2016aoc, vanHameren:2016ftb, Kovchegov:2015pbl, Kovchegov:2018znm, Kovchegov:2017lsr, Kovchegov:2021iyc, 
Chirilli:2021lif, Cougoulic:2022gbk, Kovchegov:2022kyy, Kovchegov:2024aus, Borden:2024bxa, Kovchegov:2025gcg}, one studies the small-$x$ asymptotics of GPDs by generalizing the non-local operator to that of GTMDs, allowing for an additional separation of the two field operators in the transverse direction, which will lead to an expression involving a dipole scattering amplitude. In this work, we will focus on the GPDs for unpolarized partons in unpolarized hadrons, with the relevant unpolarized dipole gluon GTMD $F_{1,1}^g$ defined by \cite{Bhattacharya:2018lgm}
\begin{align}\label{glueGTMD}
      &\frac{1}{2} \sum_S \int \frac{\dd{z^-}\dd[2]{z_{\perp}}}{(2\pi)^3} \, e^{ix \bar{P}^+ z^- - i \un{k} \vdot \un{z}} \, \delta^{ij}    \\
      &\times \, \bra{P', S} 2 \tr \Big[ F^{+i} \left(-\tfrac{z}{2} \right) {V}^{[+]} \left[ - \tfrac{z}{2} , \tfrac{z}{2} \right] F^{+j} \left(\tfrac{z}{2} \right) {V}^{[-]} \left[ \tfrac{z}{2} , - \tfrac{z}{2} \right] \Big] \ket{P, S} \notag \\
      &= \frac{1}{2} \sum_S \frac{\bar{P}^+}{2 M} \, \bar{u} (P',S) \, F_{1,1}^g (x, \xi, \un{\Delta}, \un{k}) \, u (P,S)  ,  \notag
\end{align}
where now $z^\mu = (0^+, z^-, {\un z})$, the trace and field strength operators are in the fundamental representation, $u(P,S)$ and $\bar{u} (P', S)$ are on shell spinors for the hadron of mass $M$, $\Delta = P' - P$ is the momentum transferred to the hadron, $x$ is the average longitudinal momentum fraction of the gluons, and $\xi = - \Delta^+/2 \bar{P}^+$. Note that the transverse momentum transfer $\un{\Delta}$ and average parton transverse momentum $\un{k}$ vectors are shown as arguments of GTMDs for compactness; the GTMDs actually depend on $k_T^2, \Delta_T^2$ and $\un k \cdot \un \Delta$. Similarly, we suppress the renormalization scale dependence in the arguments of GTMDs. 
In \eq{glueGTMD}, the gauge links 
\begin{align}
    V^{[\pm]} [x_f, x_i] = V_{\un{x}_f} [x^-_f, \pm \infty^-] V_{\pm \infty^-} [\un{x}_f,\un{x}_i] V_{\un{x}_i} [\pm \infty^-, x_i^-]
\end{align}
are fundamental-representation staples in the positive and negative light-cone minus directions with $V_{\pm \infty^-} [\un{x}_f,\un{x}_i]$ the transverse gauge links at infinities. 

From \eq{glueGTMD}, we see that the unpolarized gluon GTMD is
\begin{align}\label{glueGTMD111}
& F_{1,1}^g (x, \xi, \un{\Delta}, \un{k}) =  \sqrt{1-\xi^2} \sum_S \int \frac{\dd{z}^- \dd[2]{z_{\perp}} \, e^{ix \bar{P}^+ z^- - i \un{k} \vdot \un{z}}}{(2 \pi)^3 \, 2 \, \bar{P}^+}  \\
    & \times  \, \bra{P', S} 2 \tr \Big[ F^{+i} \left(-\tfrac{z}{2} \right) {V}^{[+]} \left[ - \tfrac{z}{2} , \tfrac{z}{2} \right] F^{+i} \left(\tfrac{z}{2} \right) {V}^{[-]} \left[ \tfrac{z}{2} , - \tfrac{z}{2} \right] \Big] \ket{P, S} .  \notag 
\end{align}
We will now simplify the expression \eqref{glueGTMD111} for the gluon GTMD at small $x$ and small $\xi$ by applying the Light Cone Operator Treatment (LCOT) formalism developed in \cite{Kovchegov:2015pbl,Kovchegov:2017lsr, Kovchegov:2018znm, Kovchegov:2018zeq, Kovchegov:2021iyc}, assuming that one does not choose the light-cone gauge of the hadron so that we can neglect the transverse links at infinities \cite{Collins:1992kk, Collins:2002kn, Brodsky:2002rv, Kovchegov:1996ty, Jalilian-Marian:1997xn, Kovchegov:1997ke, Kovchegov:1998bi, Brodsky:2002ue, Belitsky:2002sm, Chirilli:2015fza}. For the non-forward matrix element at hand, we write at $|x|, |\xi|\ll 1$ (see Appendix~A of \cite{Kovchegov:2019rrz})
\begin{align}\label{sat_ave}
 &\frac{1}{2} \sum_S \bra{P', S} \hat{\cal O} \left( - \alpha \, {x}_{10}, (1-\alpha) \, {x}_{10} \right) \ket{P, S} \\
 & = 2 \bar{P}^+ \, \int db^- \, d^2 b_\perp \, e^{- i \Delta^+ \, b^- + i {\un \Delta} \cdot {\un b}} \, \left\langle \hat{\cal O} \left( {x}_0, {x}_1 \right) \right\rangle \notag . 
\end{align}
Here $\hat{\cal O} \left( {x}_0, {x}_1 \right)$ is some two-point operator, as in \eq{glueGTMD111}, which depends on two positions $x_p^\mu = (0^+, x_p^-, {\un x}_p)$ with $p=0,1$. The impact parameter can be defined by 
\begin{align}
    b = \alpha \, x_1 + (1-\alpha) \, x_0
\end{align}
with $\alpha$ an arbitrary real number while $x_{10} = x_1 - x_0$. Angle brackets denote the unpolarized small-$x$/saturation averaging \cite{McLerran:1993ka, McLerran:1993ni, McLerran:1994vd, Kovchegov:1996ty, Jalilian-Marian:1997dw}: \eq{sat_ave} can be thought of as the definition of saturation averaging in the non-forward case. We note that the expression \eqref{glueGTMD111} for the gluon GTMD corresponds to $\alpha = 1/2$ in \eq{sat_ave}: however, \eq{glueGTMD111} (and the definition \eqref{glueGTMD}) can be rewritten for any other value of $\alpha$ by replacing $z/2 \to (1-\alpha) \, z$ and $-z/2 \to - \alpha \, z$ in them (see \cite{Hatta:2017cte} for a related discussion). Employing \eq{sat_ave} with $\alpha = 1/2$, and expanding in $x$ and $\xi$ to the lowest non-trivial order (cf.~\cite{Hatta:2016aoc, Kovchegov:2017lsr, Cougoulic:2022gbk} for the helicity-dependent PDFs case) using
\begin{align}\label{x-exp}
    & \int\limits_{-\infty}^\infty \dd{x^-} e^{i x  \bar{P}^+ \, x^-} V_{\un x} [\infty, x^-] \,
    F^{+i} (0^+, x^- , {\un x}) \, V_{\un x} [x^-, - \infty]   \\ 
& = - \int\limits_{-\infty}^\infty \dd{x^-} e^{i x  \bar{P}^+ \, x^-}  V_{\un x} [\infty, x^-] 
    (\pd^i A^+ + i x \bar{P}^+ \, A^i)  V_{\un x} [x^-, - \infty] \notag \\
    & = \frac{i}{g_s} \, \pd^i V_{\un x} + {\cal O} (x) \notag
\end{align}
with $x \to x + \xi$ (and the Hermitian conjugate of \eq{x-exp} with $x \to x - \xi$), we obtain (cf. \cite{Hatta:2022bxn})
\begin{align}\label{glueGTMD11}
    &F_{1,1}^g (|x| \ll 1, |\xi| \ll 1, \un{\Delta}, \un{k}) = - \frac{N_c}{8 \pi^4 \, \as}  \\
    &\times \int \dd[2]{x}_{1 \, \perp} \dd[2]{x}_{0 \, \perp} \,  e^{- i \un{k} \vdot \un{x}_{10} + i \un{\Delta} \vdot {\un b}} \, \left( \un \nabla_{{x}_1} \vdot  \un \nabla_{{x}_0} \right)   \left[ N_{10} (Y) - i \, {\cal O}_{10} (Y) \right] , \notag
\end{align}
where we have defined the unpolarized C-even dipole scattering amplitude \cite{Balitsky:1995ub, Kovchegov:1999yj}
\begin{align}
    N_{10} (Y) \equiv 1 - \frac{1}{2 N_c} \, \left\langle \tr \left[ V_{\un{x}_1} V_{\un{x}_0}^{\dagger} \right] + \tr \left[ V_{\un{x}_0} V_{\un{x}_1}^{\dagger} \right]  \right\rangle (Y)
\end{align}
along with the C-odd odderon amplitude \cite{Kovchegov:2003dm, Hatta:2005as} (see also \cite{Kovchegov:2012ga})
\begin{align}
    {\cal O}_{10} (Y) \equiv \frac{1}{2 i N_c} \, \left\langle \tr \left[ V_{\un{x}_1} V_{\un{x}_0}^{\dagger} \right] - \tr \left[ V_{\un{x}_0} V_{\un{x}_1}^{\dagger} \right]  \right\rangle (Y)
\end{align}
with the infinite fundamental light-cone Wilson lines denoted by $V_{\un{x}} = V_{\un{x}} [\infty^-, -\infty^-]$ and $Y$ the rapidity of the dipole with respect to the target. We stress again that \eq{glueGTMD11} is derived here for $\un b = (\un{x}_1 + \un{x}_0)/2$ in it, but applies for any definition of the impact parameter $\un b = \alpha \un x_1 + (1-\alpha) \un x_0$, as long as the definition \eqref{glueGTMD} is modified accordingly. The latter modification would, indeed, change the gluon GTMD $F_{1,1}^g$.

The unpolarized gluon GPD is given by \cite{Boussarie:2023izj,delRio:2024vvq}
\begin{align}\label{gGPD}
    & H^g (|x| \ll 1, |\xi| \ll 1, t, Q^2) = \int\limits^{Q^2}  \dd[2]{k}_{\perp}  F_{1,1}^g (|x| \ll 1, |\xi| \ll 1, \un{\Delta}, \un{k}) \notag \\
    & = - \frac{N_c}{2 \pi^2 \, \as} \,  \int \dd[2]{b}_{\perp} \,  e^{i \un{\Delta} \vdot {\un b}} \, \left[ \left( \nabla_{\un{x}_1} \vdot  \nabla_{\un{x}_0} \right)   N_{10} (Y) \right]_{x_{10}^2 = 1/Q^2} , 
\end{align}
where we assume that $Q^2$ is very large.
As the odderon contribution to the gluon GTMD \eqref{glueGTMD11} is odd under $\un k \to - \un k$, it does not contribute to the gluon GPD in \eq{gGPD}.

The unpolarized quark GTMD is defined by \cite{Meissner:2009ww}
\begin{align}\label{qGTMD}
    &\frac{1}{2} \sum_S \int \frac{\dd{z^-}\dd[2]{z_{\perp}}}{2 (2\pi)^3} e^{ix \bar{P}^+ z^- - i \un{k} \vdot \un{z}} \\
    &\times \, \bra{P', S} \bar{\psi} \left(-\tfrac{z}{2} \right) V^{[+]} \left[ -\tfrac{z}{2},\tfrac{z}{2} \right] \gamma^+ \psi \left( \tfrac{z}{2} \right) \ket{P, S} \notag  \\
      &= \frac{1}{4 M} \sum_S \bar{u} (P',S) \, F_{1,1}^q (x, \xi, \un{\Delta}, \un{k}) \, u(P,S) . \notag
\end{align}
We chose the future-pointing Wilson line staple, as used in semi-inclusive deep inelastic scattering (SIDIS). 
To simplify the quark GTMD at $|x|, |\xi| \ll 1$, we start with the definition \eqref{qGTMD}, employ \eq{sat_ave}
and insert a complete set of states to obtain (while approximating $\sqrt{1-\xi^2} \approx 1$ at $|\xi| \ll 1$)
\begin{align}\label{qGTMD_smallx}
    F_{1,1}^q &(|x| \ll 1, |\xi| \ll 1, \un{\Delta}, \un{k}) = \frac{\bar{P}^+}{(2 \pi)^3}  \int \dd[2]{x}_{1 \, \perp} \dd{x}_1^- \dd[2]{x}_{0 \, \perp} \dd{x}_0^- \notag   \\
    &\times e^{i (x + \xi) \bar{P}^+ x_1^- - i (x - \xi) \bar{P}^+ x_0^-} e^{ - i \un{k} \vdot \un{x}_{10}  + i \un{\Delta} \vdot (\un{x}_1 + \un{x}_0)/2 } \, \left( \gamma^+ \right)_{\alpha\beta} \\
    &\times \sum_X\Big\langle  \bar{\psi}_{\alpha} (x_0) V_{\un{x}_0} [x_0^-, \infty] \ket{X} \bra{X} | V_{\un{x}_1} [\infty, x_1^-] \psi_{\beta} (x_1) \Big\rangle  \notag
\end{align}
with the implied summation over the quark spinor indices $\alpha, \beta$.
The complete set of final states allows us to calculate this non time-ordered matrix element \eqref{qGTMD} diagrammatically, whereas one could also work with a time-ordered matrix element (see \cite{Diehl:1998sm} for a discussion on the difference between time-ordered and non time-ordered matrix elements in the collinear framework) and then apply diagrammatic methods without inserting the complete set of states (as in \cite{Bhattacharya:2025fnz}). Following the calculation detailed in \cite{Kovchegov:2021iyc, Kovchegov:2018znm, Borden:2024bxa}, one obtains, for massless quarks (cf.~\cite{Bhattacharya:2025fnz}),
\begin{align}\label{q_GTMD_1}
   &F_{1,1}^q (|x| \ll 1, |\xi| \ll 1, \un{\Delta}, \un{k})= - \frac{2 N_c \, s}{(2\pi)^4} \int \dd[2]{x}_{1 \, \perp} \dd[2]{x}_{0 \, \perp} \\
   &\times \int\frac{\dd[2]{k}_{1 \, \perp}}{(2\pi)^2} \, e^{-i(\un{k}_1 + \un{k})\vdot \un{x}_{10} + i \un{\Delta} \vdot (\un{x}_1 + \un{x}_0)/2} \, \int\limits_{\Lambda^2/s}^1 \dd{z}  \, \left[ N_{10} (Y) - i \, {\cal O}_{10} (Y) \right] \notag \\
   &\times  \left\{ \frac{ \un{k}_1 \vdot \left( \un{k} + \thalf \un \Delta \right) }{\left[ (x+\xi) \, z\, s + \un{k}_1^2 - i \epsilon \right] \left[ (x-\xi) \, z \, s + \left( \un{k} + \thalf \un \Delta \right)^2 - i \epsilon \right]} \right. \notag \\
   & + \frac{ \un{k}_1 \vdot \left( \un{k} - \thalf \un \Delta \right) }{\left[ (x+\xi) \, z\, s + \left( \un{k} - \thalf \un \Delta \right)^2 + i \epsilon \right] \left[ (x-\xi) \, z \, s + \un{k}_1^2 + i \epsilon \right]} \notag \\
   & \left. + \frac{\un{k}_1^2}{\left[ (x+\xi) \, z\, s + \un{k}_1^2 - i \epsilon \right] \left[ (x-\xi) \, z \, s + {\un k}_1^2 + i \epsilon \right]}  \right\}   \notag
\end{align}
with $\Lambda$ an infrared (IR) scale describing the target. 
Note also that we had to add the diagram D in the notation of \cite{Kovchegov:2021iyc}, which was not included in the calculation in \cite{Kovchegov:2021iyc} because the calculation there was for a target spin-dependent distribution, while diagram D brings in the non-interaction term where none of the Wilson lines pass through the shock wave and is, therefore, independent of the hadron spin. The variable $s \approx 2 \bar{P}^+ q^-$ in \eq{q_GTMD_1} denotes the center-of-mass energy squared for the projectile dipole--target system, with $z$ the fraction of the projectile ``minus" momentum $q^-$ carried by the anti-quark in the dipole: while the variable $z$ (or $z s$) can be integrated out by assuming a relatively weak $z$-dependence in $N_{10}$ and ${\cal O}_{10}$ (see \cite{Kovchegov:2015zha, Kovchegov:2021iyc, Bhattacharya:2025fnz}), this simplification is not our goal here. We do note that, as follows from the $z$-integral in \eq{q_GTMD_1}, the typical $z$-value is order-1, such that it does not significantly affect the value of rapidity $Y$ in the argument of the dipole amplitude (as $\ln (zs) \approx \ln (s)$ for $z = {\cal O} (1)$, the rapidity $Y$, which usually depends on $\ln (zs)$, can be thought of as dependent on $\ln (s)$ instead). Since $N_{10} (Y) = N_{01} (Y)$ and ${\cal O}_{10} (Y) = - {\cal O}_{01} (Y)$, one can readily see that our expression \eqref{q_GTMD_1} for quark GTMD satisfies $\left[ F_{1,1}^q (x, \xi, \un{\Delta}, \un{k}) \right]^* = F_{1,1}^q (x, - \xi, - \un{\Delta}, \un{k})$, in agreement with Eq.~(3.18) in \cite{Meissner:2009ww} following from Hermiticity, as long as $N_{10} (Y)$ and ${\cal O}_{10} (Y)$ are even functions of $\xi$.

The corresponding unpolarized quark GPD at $|x|, |\xi| \ll 1$ is given by
\begin{align}\label{qGPD}
    H^q &(|x|\ll 1, |\xi| \ll 1, t, Q^2) = \int\limits^{Q^2}  \dd[2]{k}_{\perp} F_{1,1}^q (|x| \ll 1, |\xi| \ll 1, \un{\Delta}, \un{k}) . 
\end{align}
It appears that, unlike the $H^g$ case, the odderon contribution does not vanish in $H^q$, but only in the Efremov-Radyushkin-Brodsky-Lepage (ERBL) \cite{Efremov:1978rn,Lepage:1979zb} region, $|x| < |\xi|$.

We have obtained expressions \eqref{glueGTMD11} and \eqref{q_GTMD_1} for both the gluon and quark unpolarized GTMDs at small $x$ and $\xi$ in terms of the impact parameter-dependent dipole scattering amplitudes; these GTMDs give their corresponding GPDs after integrating over the parton transverse momentum (see Eqs.~\eqref{gGPD} and \eqref{qGPD}). However, the rapidity dependence of the dipole amplitude in these formulas is not yet fixed. Usually, for PDFs and TMDs at small $x$, one takes $Y = \ln (1/x)$. The question we would like to address in this work is what $Y$ should one use in Eqs.~\eqref{glueGTMD11}, \eqref{gGPD}, \eqref{q_GTMD_1} and \eqref{qGPD}? There is a non-perturbative $(x,\xi)$ dependence in the initial condition of the off-forward matrix element defining the GTMDs and GPDs that we will not address here. Instead, in the following Section we will examine the role of $\xi$ in the evolution of the dipole amplitude and establish a prescription for including effects of non-zero skewness into the dipole scattering amplitude(s) resulting from the full non-linear evolution equation. We will find that the lifetime ordering required for resumming the leading logarithms will give us an effective rapidity $Y$ for the evolved dipole amplitudes $N_{10} (Y)$ and ${\cal O}_{10} (Y)$ which will depend on the skewness.

%%%%%%%%%%%%%%%%%%%%%%%%%%%%%%

\section{Skewness-dependent dipole amplitudes resulting from non-linear evolution}

\label{sec:skew}

%%%%%%%%%%%%%%%%%%%%%%%%%%%%%%%%%%%%%%%%%%%%%%%%%%%%%%%%%%%%%%%%%%%%%%%%%%%%%%%%%%%%%%
\begin{figure}[ht]
\centering
\includegraphics[width= \linewidth]{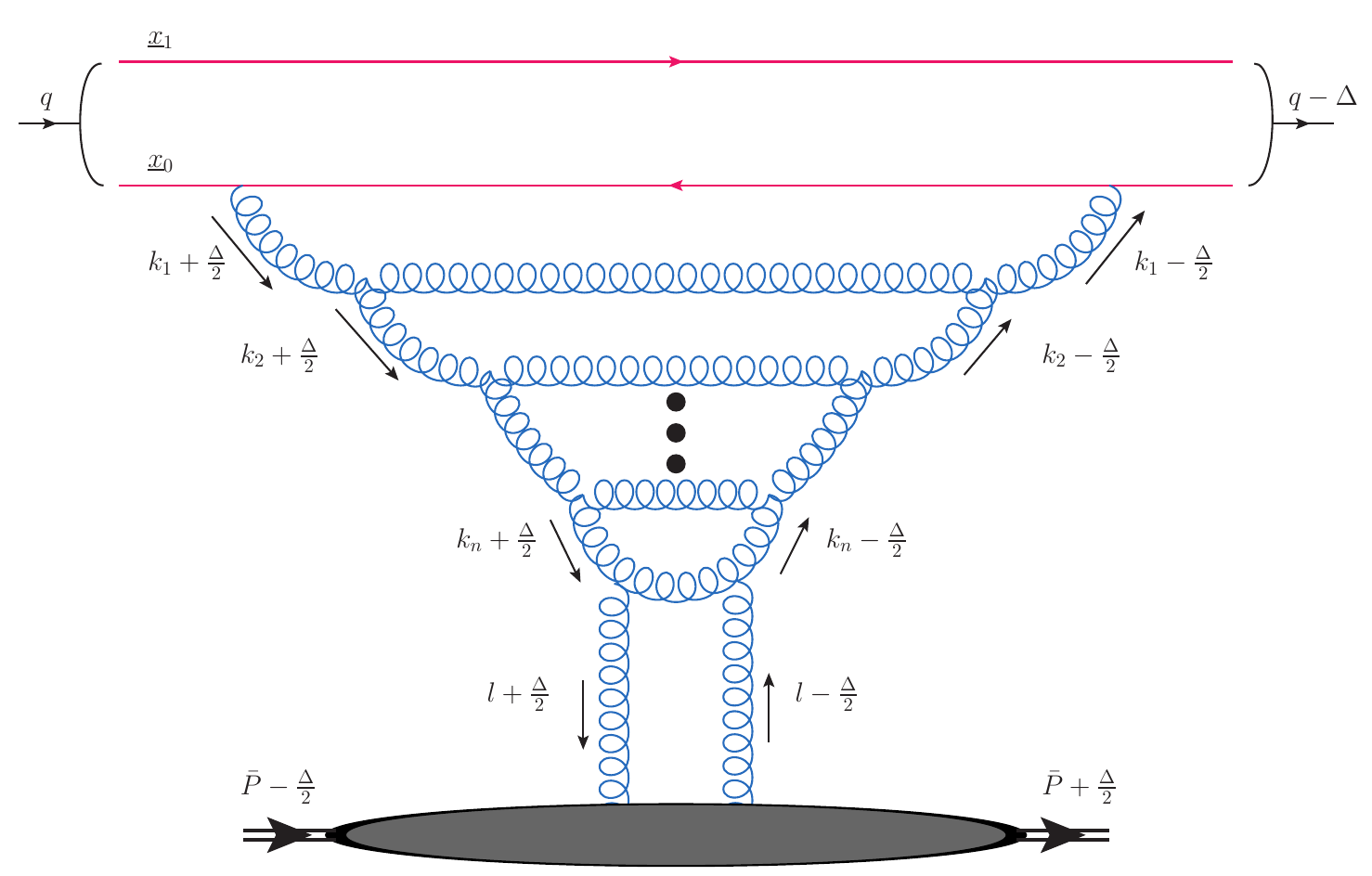}
\caption{A ladder diagram contributing to the small-$x$ evolution of a quark and anti-quark dipole in the general non-forward case for an elastic scattering process. }
\label{FIG:skew_lad}
\end{figure}
%%%%%%%%%%%%%%%%%%%%%%%%%%%%%%%%%%%%%%%%%%%%%%%%%%%%%%%%%%%%%%%%%%%%%%%%%%%%%%%%%%%%%%

The dipole amplitudes $N_{10}$ and ${\cal O}_{10}$ resum diagrams for the exchange of arbitrarily many eikonal gluons between a quark and anti-quark color dipole and a target. Within the strict eikonal (LLA) limit, one would ignore the longitudinal momentum transfer, allowing only for transverse momentum transfer $\un \Delta$ which is Fourier conjugate to the impact parameter of the dipole, as is shown explicitly in Eqs.~\eqref{glueGTMD11} and \eqref{q_GTMD_1}. In order to incorporate skewness dependence, we need to consider a fully non-forward scattering where we keep track of the full (transverse and longitudinal) momentum transfer in the diagrams. Such a calculation was performed in the case of the BFKL \cite{Kuraev:1977fs,Balitsky:1978ic} ladder diagrams (with a running coupling prescription) in \cite{Bartels:1981jh}, where transverse and longitudinal momentum transfers were explicitly kept in the diagrams. Here we will show how to incorporate the correction for longitudinal momentum transfer for arbitrary diagrams contributing to the unpolarized small-$x$ evolution, working in the shock wave/saturation framework and thus making use of impact-parameter-dependent dipole scattering to handle the dependence on transverse momentum transfer. Including general diagrams will allow us to derive a prescription for including the longitudinal momentum transfer not only into the  dipole amplitude $N_{10}$ resulting from the linear evolution %regime 
described by the BFKL equation, but also into $N_{10}$ in the non-linear/saturation regime, where it is described by the Balitsky-Kovchegov (BK) \cite{Balitsky:1995ub, Balitsky:1998ya, Kovchegov:1999yj, Kovchegov:1999ua} and Jalilian-Marian-Iancu-McLerran-Weigert-Leodinov-Kovner (JIMWLK) \cite{Jalilian-Marian:1997jx,Jalilian-Marian:1997gr,Jalilian-Marian:1997dw,Iancu:2001ad,Iancu:2000hn} equations. %for the anplitude $N_{10}$. 
The small-$x$ odderon amplitude ${\cal O}_{10}$, whose evolution can be found in \cite{Bartels:1999yt,Kovchegov:2003dm,Hatta:2005as} (see also \cite{Kovchegov:2012rz, Janik:1998xj, Caron-Huot:2013fea,Bartels:2013yga, Brower:2008cy,Avsar:2009hc,Brower:2014wha}), will also be subject to the same prescription for including $\xi \neq 0$ effects. 

Let us start by discussing the specific case of ladder diagrams contributing to a non-forward elastic process, as shown in \fig{FIG:skew_lad}: note that our discussion here will be generalized to the multi-ladder diagrams. Imagine that a projectile comes in with momentum $q^\mu = (- Q^2/(2 q^-), q^-, {\un 0})$, where $q^-$ is large. While the projectile is not necessary for the GTMDs and GPDs at hand, it is useful for imposing the conditions below on the kinematics of the gluons in the cascade. The projectile produces a dipole with the quark at position $\un x_1$ and the anti-quark at position $\un x_0$, which, through small-$x$ evolution, source a gluon cascade, with the momenta of the gluons in the $i$th step of evolution (the ``rails" of the ladder) equal to 
\begin{align}
    k_i^\mu = \left( \frac{\left({\un k}_{i} \pm \tfrac{\un \Delta}{2} \right)^2}{2 (k_i^- \pm \tfrac{\Delta^-}{2})} , k_i^- \pm \frac{\Delta^-}{2} , {\un k}_{i} \pm \frac{\un \Delta}{2} \right).
\end{align} 
We are assuming a light-cone perturbation theory (LCPT) calculation \cite{Lepage:1980fj, Brodsky:1997de}. Below, we will also employ the ``minus" light cone momentum fractions $z_i = k_i^- / q^-$ and $\zeta = \Delta^- / 2 q^-$. 

The gluon cascade generating small-$x$ evolution must obey the following conditions, in order to be leading logarithmic (see, e.g., \cite{Kovchegov:2016zex, Cougoulic:2019aja} for a detailed derivation). First of all, the light-cone ``minus" momenta have to be ordered, 
\begin{align}\label{k_minus_ordering}
    q^- \gg k_1^- \pm \frac{\Delta^-}{2} \gg k_2^- \pm \frac{\Delta^-}{2} \gg \ldots \gg k_n^- \pm \frac{\Delta^-}{2} . 
\end{align}
(Note also that $k_i^- \pm \tfrac{\Delta^-}{2} >0$ by the LCPT rules.) The light-cone lifetimes of the gluons should also be ordered, leading to the ordering of the light-cone energy denominators for the successive emissions according to
\begin{align}\label{lifetime1}
    \frac{Q^2}{2q^-}, |\Delta^+| \ll \frac{(\un{k}_1 \pm \un{\Delta}/2)^2}{2(k_1^- \pm \Delta^-/2)}  \ll \frac{(\un{k}_2 \pm \un{\Delta}/2)^2}{2(k_2^- \pm \Delta^-/2)}  \ll \ldots \\
    \ll \frac{(\un{k}_{n-1} \pm \un{\Delta}/2)^2}{2(k_{n-1}^- \pm \Delta^-/2)}  \ll \frac{(\un{k}_n \pm \un{\Delta}/2)^2}{2(k_n^- \pm \Delta^-/2)}  . \notag
\end{align} 
Finally, the transverse momenta must be of the same order
\begin{align}\label{t_mom_ord}
    \left| \un{k}_1 \pm \frac{\un \Delta}{2}  \right| \sim \left| \un{k}_2 \pm \frac{\un \Delta}{2}  \right| \sim ...  \sim \left| \un{k}_n \pm \frac{\un \Delta}{2}  \right| ,
\end{align}
and not too different from the hard scale $Q$.

The ordering in \eq{k_minus_ordering} cannot be realized if $k_i^-$ momenta are comparable to or smaller (in magnitude) than $|\Delta^-|$. Which means, we need to have $k_i^- \gg \Delta^-$. We note that $\Delta^- / q^-$ is suppressed by an inverse power of energy $s$, such that we can neglect $\Delta^-$ compared to $k_i^-$ in the gluon lines far from the bottom of the ladder. However, as further emissions drive the $n$th ``minus" momentum $k_n^-$ down toward its minimum value, one can no longer neglect $\Delta^-$ and eventually one would find either $k_n^- + (\Delta^-/2) < 0$ or $k_n^- - (\Delta^-/2) < 0$, which violates the ordering \eqref{k_minus_ordering} and the forward momentum flow of the on-shell internal lines in LCPT. The minimum value of the ``minus" momentum fractions $z_n$ set by the shock wave is $\Lambda^2/s$, for a scale $\Lambda$ characterizing the non-perturbative dynamics of the target. Thus, if $\Lambda^2/s < |\zeta|$, the forward momentum flow of the gluon cascade diagrams requires that we only evolve down to $z_n = |\zeta|$, while for $\Lambda^2/s > |\zeta|$ we can evolve all the way down to the shock wave cutoff $\Lambda^2/s$. This means that the full ordering of the longitudinal momentum fractions is
\begin{align}\label{beta-ordering}
    z_1 \gg z_2 \gg ... \gg z_{n-1} \gg z_n \gg \max \left\{ \Lambda^2/s, |\zeta| \right\} .
\end{align}
The diagram containing $n$ steps of evolution ($n$ $s$-channel gluons) will give us a leading logarithmic contribution given by
\begin{align}
    \alpha_s^{n} \int\limits_{z_{\min}}^{z} \frac{\dd{z_1}}{z_1} \int\limits_{z_{\min}}^{z_2} \frac{\dd{z_2}}{z_2}\dots\int\limits_{z_{\min}}^{z_{n-1}} \frac{\dd{z_{n}}}{z_{n}}
    = \frac{1}{n!}\alpha_s^{n} \ln^{n} \Bigg(\frac{z}{z_{\min}} \Bigg) 
\end{align}
with $z_{\min} = \max \left\{ \Lambda^2/s, |\zeta| \right\}$.

Here we assume that the total momentum transfer $|t| \approx \un{\Delta}^2$ is small, as in the usual collinear factorization kinematics, where we take $t$ to be a of the same order as the IR scale of the target: $|t| \sim \Lambda^2 \ll Q^2$. In these kinematics, the shock wave cutoff from the forward limit will generically be of the same order as the $\zeta$-term, since $|\zeta| \sim \Delta_\perp^2/s \approx |t|/s \sim \Lambda^2/s$. Thus, we can neglect the difference between $|\zeta|$ and $\Lambda^2/s$ in $\ln z_{min}$.

%%%%%%%%%%%%%%%%%%%%%%%%%%%%%%%%%%%%%%%%%%%%%%%%%%%%%%%%%%%%%%%%%%%%%%%%%%%%%%%%%%%%%%
\begin{figure}[ht]
\centering
\includegraphics[width= \linewidth]{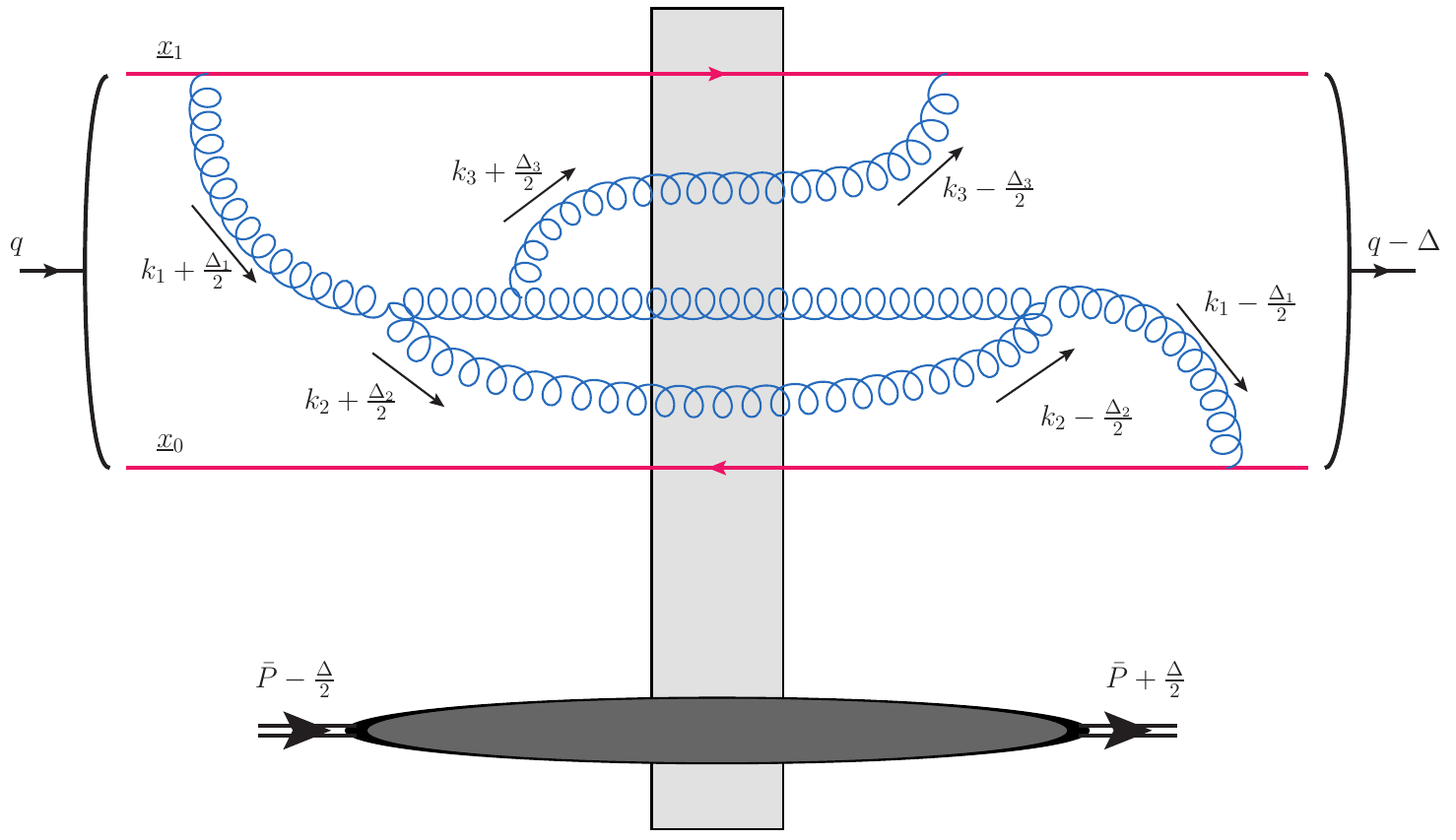}
\caption{A gluon cascade in the non-forward elastic scattering case. The rectangle represents the shock wave, comprising subsequent evolution and interaction with the target.}
\label{FIG:skew_cascade}
\end{figure}
%%%%%%%%%%%%%%%%%%%%%%%%%%%%%%%%%%%%%%%%%%%%%%%%%%%%%%%%%%%%%%%%%%%%%%%%%%%%%%%%%%%%%%

In the dipole picture \cite{Mueller:1994rr,Mueller:1994jq,Mueller:1995gb}, even the linear BFKL evolution is not limited to ladder diagrams. The non-linear evolution \cite{Balitsky:1995ub, Balitsky:1998ya, Kovchegov:1999yj, Kovchegov:1999ua,Jalilian-Marian:1997jx,Jalilian-Marian:1997gr,Jalilian-Marian:1997dw,Iancu:2001ad,Iancu:2000hn} requires gluon cascades to continue in different dipoles. This is illustrated in \fig{FIG:skew_cascade}, where the original dipole $10$ is split into two by the emission of the gluon $k_1$, with the subsequent emissions of gluons $k_2$ and $k_3$ taking place in each of the two new dipoles. The shock wave is represented by the shaded rectangle. Note that now the momentum transfer is different in each step of the evolution, such that the momenta of the gluons in the $i$the step of evolution can be labeled as
\begin{align}
    k_i^\mu = \left( \frac{\left({\un k}_{i} \pm \tfrac{\un \Delta_i}{2} \right)^2}{2 (k_i^- \pm \tfrac{\Delta^-_i}{2})} , k_i^- \pm \frac{\Delta^-_i}{2} , {\un k}_{i} \pm \frac{\un \Delta_i}{2} \right).
\end{align} 
Once again, we argue that the ``minus" momentum ordering and/or the $k^- >0$ condition in LCPT require that $k_i^- \gg \Delta^-$, which results in \eq{beta-ordering}. (A gluon with $k_i^- \lesssim \Delta^-$ has the $x^-$-lifetime shorter than the width of the shock wave, and is, therefore, outside of our eikonal approximation.) Equations \eqref{lifetime1} and \eqref{t_mom_ord} become
\begin{align}\label{lifetime2}
    \frac{Q^2}{2q^-}, |\Delta^+| \ll \frac{(\un{k}_1 \pm \un{\Delta}_1/2)^2}{2(k_1^- \pm \Delta_1^-/2)}  \ll \frac{(\un{k}_2 \pm \un{\Delta}_2/2)^2}{2(k_2^- \pm \Delta_2^-/2)}  \ll \ldots \\
    \ll \frac{(\un{k}_{n-1} \pm \un{\Delta}_{n-1}/2)^2}{2(k_{n-1}^- \pm \Delta_{n-1}^-/2)}  \ll \frac{(\un{k}_n \pm \un{\Delta}_n/2)^2}{2(k_n^- \pm \Delta_n^-/2)}   \notag
\end{align} 
and
\begin{align}\label{t_mom_ord2}
    Q \sim \left| \un{k}_1 \pm \frac{\un \Delta_1}{2}  \right| \sim \left| \un{k}_2 \pm \frac{\un \Delta_2}{2}  \right| \sim ...  \sim \left| \un{k}_n \pm \frac{\un \Delta_n}{2}  \right| .
\end{align}

Above, we have demonstrated how the longitudinal (``minus") momentum transfer affects the cutoff on the ``minus" longitudinal momentum fraction for emitted gluons in the cascade. As skewness is a plus momentum fraction, we need to devise a similar argument for the ``plus" momenta. Thus, we need to show how the conditions \eqref{lifetime2} arising from the lifetime ordering, which is equivalent to the ``plus" momentum ordering, lead to a dependence of the dipole amplitudes on skewness. In the language of LCPT the leading logarithmic contribution comes from diagrams where the energy denominator is dominated by the latest gluon emission. On the left side of the shock wave this means 
\begin{align}
    \frac{1}{(k_i + \Delta_i/2)^+ - q^+} = \frac{1}{\frac{(\un{k}_i + \un{\Delta}_i/2)^2}{2(k_i^- + \Delta_i^-/2)} - q^+} \approx  \frac{1}{\frac{(\un{k}_i + \un{\Delta}_i/2)^2}{2(k_i^- + \Delta_i^-/2)} } ,
\end{align}
implying 
\begin{align}
    \frac{(\un{k}_i + \un{\Delta}_i/2)^2}{2(k_i^- + \Delta_i^-/2)} \gg q^+ .
\end{align}
This condition is generically true for small-$x$ evolution as $q^+ \sim Q^2/q^-$ with $q^- \gg k_i^- + \Delta_i^-/2$: hence, this is not a new constraint. On the right side of the shock wave we have 
\begin{align}
    \frac{1}{\frac{(\un{k}_i - \un{\Delta}_i/2)^2}{2(k_i^- - \Delta_i^-/2)} - (q^+ + \Delta^+)} \approx  \frac{1}{\frac{(\un{k}_i - \un{\Delta}_i/2)^2}{2(k_i^- - \Delta_i^-/2)} } ,
\end{align}
implying 
\begin{align}
    \frac{(\un{k}_i - \un{\Delta}_i/2)^2}{2(k_i^- - \Delta_i^-/2)} \gg q^+, |\Delta^+ | .
\end{align}
The inequality is true for $q^+$ as long as the lifetime ordering we established in the ``minus" momenta is satisfied in the forward limit, while the relation with $\Delta^+$ implies that
\begin{align}
     \frac{(\un{k}_i - \un{\Delta}_i/2)^2}{2(k_i^- - \Delta_i^-/2)} \gg 2 |\xi| \bar{P}^+ .
\end{align}
Using the $k_i^-=z_i q^- \gg \Delta_i^-/2$ condition we have
\begin{align}\label{lto}
    \frac{(\un{k}_i - \un{\Delta}_i/2)^2}{z_is} \gg 2 |\xi| .
\end{align}
The expression on the left of \eq{lto} is proportional to the inverse of the $x^-$-lifetime of the $i$th gluon. Therefore, it is smallest for the gluon with the longest lifetime, that is, for the first gluon in the cascade. This means the strongest version of the constraint is for $i=1$,
\begin{align}
    \frac{(\un{k}_1 - \un{\Delta}_1 /2)^2}{z_1 s} \gg 2 |\xi| ,
\end{align}
where the initial dipole size $x_{10} \sim 1/Q$ sets the typical momentum for the first emitted gluon as $|\un{k}_1 - \un{\Delta}_1 /2| \sim 1/{x_{10}} \sim Q$, yielding
\begin{align}
    z_1 \ll \min \left\{ 1, \, \frac{x}{2 |\xi|} \right\} ,
\end{align}
where we have defined $x = Q^2/s$ and employed the fact that $z_1 \ll 1$ in the LLA. The integral over $z_1$ at LLA in the first step of evolution, therefore, gives
\begin{align}\label{zint}
    \int\limits_{\frac{1}{s x_{10}^2}}^{\min \left\{ 1, \frac{x}{2 |\xi|} \right\}} \frac{d z_1}{z_1} \approx \int\limits_{x}^{\min \left\{ 1, \frac{x}{2 |\xi|} \right\}} \frac{d z_1}{z_1} \approx \ln \left( \min \left\{ \frac{1}{x} , \frac{1}{|\xi|} \right\} \right) .
\end{align}
Here we have chosen the lower limit of the $z_1$ integral to be $1/s x_{10}^2$ for simplicity: our conclusion would not change for other lower limits (of the same parametric dependence on $s$). We have also neglected a factor of 2 under the logarithm with the leading logarithmic accuracy.

Since the subsequent gluon emissions obey the minus momentum ordering, $z_i \ll z_1$, the result \eqref{zint} sets the rapidity in the argument of the dipole amplitudes $N$ and $\cal O$. Thus we can write the expressions for the gluon GPD as 
\begin{align}\label{gGPD_ev}
    & H^g (|x| \ll 1, |\xi| \ll 1, t, Q^2) = - \frac{N_c}{2 \pi^2 \, \as} \int \dd[2]{b}_{\perp} \,  e^{i \un{\Delta} \vdot {\un b}} \\
    & \times \, \left[ \left( \nabla_{\un{x}_1} \vdot  \nabla_{\un{x}_0} \right)   N_{10} \left(Y = \ln \left( \min \left\{\frac{1}{|x|}, \frac{1}{|\xi|} \right\} \right) \right)  \right]_{x_{10}^2 = 1/Q^2} . \notag
\end{align}
For the quark GPD along with the gluon and quark GTMDs one has to use 
\begin{align}\label{Y}
    Y = \ln \left( \min \left\{\frac{1}{|x|}, \frac{1}{|\xi|} \right\} \right)
\end{align}
in Eqs.~\eqref{glueGTMD11}, \eqref{q_GTMD_1}, and \eqref{qGPD}. We have replaced $x \to |x|$ to account for possible negative values of the average momentum fraction $x$.

This is the main result of this work: the effect of skewness on  
the unpolarized GPDs and GTMDs at low $x$ is to make the effective rapidity of the unpolarized dipole amplitudes $N_{10} (Y)$ and ${\cal O}_{10} (Y)$ (that these distributions depend on) determined by the minimum of $1/|x|$ and $1/|\xi|$, as shown in \eq{Y}. 

%%%%%%%%%%%%%%%%%%%%%%%%%%%%

\section{Effects of skewness in high energy exclusive processes and GPDs}\label{sec:obs}

Having derived the correction for skewness in the GPDs at small-$x$ and small-$\xi$, we can see how this correction affects an observable. Let us consider a generic deep inelastic scattering (DIS)-like process where an incoming probe with virtuality $q^2 = -Q^2$ quasi-elastically scatters off a hadronic target with averaged momentum $\bar P$ and produces an outgoing particle with momentum $q' = q - \Delta$, where we label $q'^2 = Q'^2$, following the usual DIS notation. Taking the center of mass energy squared for the process to be $s = 2 \bar{P} \vdot q \approx 2 \bar{P}^+ q^-$, we define $x = Q^2/s$ and 
\begin{align}
    \xi = - \frac{\Delta^+}{2 \bar{P}^+} \approx  \frac{Q^2 + Q^{\prime \, 2}}{2s} .
\end{align}
In the case where a light-like particle is produced ($Q^{\prime \, 2} =0$, e.g., in DVCS), we have $2 \xi = x$ leading to no effect of skewness on the dipole amplitude in the LLA, since $\xi$ and $x$ are of comparable magnitude. In the case of a particle being produced with a large time-like momentum (DVMP or double DVCS) we can have $Q'^2 \gg Q^2 $ and thus $|\xi| \gg x$, such that the rapidity of the dipole amplitudes $N$ and $\cal O$ is set by $\xi$. This matches the treatment of processes such as the exclusive leptoproduction of heavy vector mesons in the literature, where we have $Q'^2 = M_h^2$ with the heavy meson mass $M_h$, thus setting the rapidity to be defined by $x$ of the parton, defined as $x_{\mathbb{P}} = 2 |\xi|$ (cf.~\cite{Mantysaari:2016jaz}). Here we have shown that this effective rapidity comes directly from a leading logarithmic analysis of the evolution of the dipole amplitude with non-zero longitudinal momentum transfer.

We can also examine the effect of the skewness correction on the phenomenological GPDs at small $x$ and $\xi$. Often GPD models and parameterizations at small-$x$ and small-$\xi$ find that GPDs and PDFs are related by an order one parameter $R$, defined (for gluons) as
\begin{align}
    R_g(x) = \frac{H_g(x,\xi=x,0)}{xf_g(x)} .
\end{align}
(For brevity, we suppress the dependence on the renormalization scale $Q^2$ in all terms here and below.)
This $R_g$ parameter is usually determined from GPD models and allows one to approximate the gluon GPD $H_g$ at small-$x$ and small-$\xi$ if the gluon PDF $f_g$ is known. Indeed, one can argue based on an expansion of the conformal moments of the GPDs at small-$\xi$ \cite{Shuvaev:1999ce,Martin:2008gqx} that this ratio is generically greater than one. Taking into account the evolution of the GPDs at small-$x$ and small-$\xi$ outside of the saturation regime, we find that a similar ratio,
\begin{align}
    R (x,\xi) = \frac{H_g(x, \xi)}{xf_g(x)} \Bigg|_{|x| \ll1, \, |\xi| \ll 1} \sim \Bigg(\frac{|x|}{\max\{|x|,|\xi| \}}\Bigg)^{\alpha_P - 1} ,
\end{align}
is less than or equal to one. (Here $\alpha_P - 1 = \frac{4 \as N_c}{\pi} \, \ln 2$ is the BFKL pomeron intercept in the LLA.) This is not in conflict with the phenomenological results: firstly because we do not account for non-perturbative effects in the initial condition of the dipole amplitude, and secondly because this asymptotic solution for the GPD only holds for $|x| \ll |\xi|$ or $|x| \gg |\xi|$ while the ratio $R_g$ usually employed in phenomenology is taken at exactly $|x| = |\xi|$. However, it is quite interesting to note that resumming the leading logarithms in $x$ and $\xi$ seems to yield a very different behavior from the resummation of leading logarithms in $Q^2$ for the ratio $R_g$, with our $R (x,\xi) \ll 1$ deep in the ERBL region, $|x| \ll |\xi|$. 

%%%%%%%%%%%%%%%%%%%%%%%%%%%%

\section{Conclusions}\label{sec:conc}

We have studied the small-$x$ and small-$\xi$ limits of the unpolarized quark and gluon GPDs and GTMDs, obtaining expressions for the unpolarized distributions in terms of the amplitudes $N$ and $\cal O$ for a color dipole scattering off a hadronic or nuclear target. In order to include the effect of skewness, we allowed for longitudinal momentum transfer in the dipole-hadron scattering process. Working to leading logarithmic accuracy, we have shown that the effect of longitudinal momentum transfer is to shift the rapidity scale to which the dipole amplitudes $N$ and $\cal O$ are evolved. This means that the dipole amplitudes should be evolved from their initial conditions at some low initial rapidity to the effective rapidity $Y = \ln ( \min \{1/|x|, 1/|\xi| \})$, with no changes to the small-$x$ evolution equation at fixed impact parameter. This conclusion holds for linear and non-linear evolution, and thus can be applied to the BFKL, BK and JIMWLK equations to study the high-energy limit of unpolarized GPDs and GTMDs. 

Including skewness in the LLA dipole scattering amplitude opens the way for the study of all leading-twist GPDs and GTMDs at small-$x$ and small-$\xi$ via the LCOT. Here we have focused only on the eikonal case, demonstrating that the prescription \eqref{Y} for skewness-dependent rapidity applies to all of the (non-forward) eikonal parton distributions at LLA. However, spin-dependent parton distributions are often zero at the eikonal order. Going beyond this approximation to study sub-eikonal distributions will likely result in order-$x$ and order-$\xi$ terms, possibly allowing one to identify not only the spin-dependent operators needed to generate polarized GPDs/GTMDs, but also the operators which will bring in explicit skewness dependence outside of the evolution logarithms (that is, order-$\xi$ terms). Such contributions may yield numerically significant corrections even though they are parametrically small, as seen in the non-perturbative skewness dependence of GPDs in high-energy scattering processes \cite{Toll:2012mb,Mantysaari:2016jaz,Cuic:2023mki,Guo:2024wxy}. Sub-eikonal corrections include operators which exchange flavor information \cite{Borden:2024bxa, Chirilli:2021lif}, so one also could expand the small-$x$ framework to study transition GPDs/GTMDs where the initial and final hadron are of different species. 

%%%%%%%%%%%%%%%%%%%%%%%%%%%%%%%%%%

\section{Acknowledgments}

The authors are grateful to Renaud Boussarie, Volodya Braun, Hervé Dutrieux, Yuxun Guo, Yoshitaka Hatta, and Feng Yuan for informative discussions. 

The work of YK and HS is supported by
the U.S. Department of Energy, Office of Science, Office of Nuclear Physics under Award Number DE-SC0004286 and within the framework of the Saturated Glue (SURGE) Topical Theory Collaboration. The work of MGS is supported by the Center for Nuclear Femtography, Southeastern Universities Research Association, Washington, D.C. and U.S. DOE Grant number DE-FG02-97ER41028, and also by the U.S. Department of Energy, Office of Science, Office of Nuclear Physics under Award Number ~DE-AC05-06OR23177 under which Jefferson Science Associates, LLC, manages and operates Jefferson Lab.

%%%%%%%%%%%%%%%%%%%%%%%%%%%%%%%%%%

%\bibliographystyle{elsarticle-num}
%\bibliography{references}

\end{document}